\documentclass[aps,prd,10pt,twocolumn,superscriptaddress,floatfix]{revtex4}
\usepackage{times}
\usepackage{amsmath,amssymb,amsfonts}
\usepackage{bm}
\usepackage{xcolor}
\usepackage{graphicx}
\usepackage{booktabs}
\usepackage{placeins}
\usepackage{orcidlink}
\usepackage{microtype}
\usepackage{xurl}
\usepackage{hyperref}
\hypersetup{
  colorlinks=true,
  linkcolor=blue,
  citecolor=blue,
  urlcolor=blue,
  filecolor=blue
}
\newcommand{\doilink}[1]{\href{https://doi.org/#1}{doi:#1}}
\newcommand{\todo}[1]{}

\begin{document}

\title{Circulation and operational sparsity of analogue Hawking radiation
in rotating acoustic horizons}

\author{Fernando M. Belchior\orcidlink{0009-0006-8675-7849}}
\email{fernandobelcks7@gmail.com}
\affiliation{Departamento de Física, Universidade Federal da Paraíba, Centro de Ciências Exatas e da Natureza, 58051-970, João Pessoa, Paraíba, Brazil}

\author{Jo\~{a}o A.A.S.\ Reis\orcidlink{0000-0002-2831-5317}}
\email{joao.reis@uesb.edu.br}
\affiliation{Departamento de Ci\^{e}ncias Exatas e Naturais, \\
Universidade Estadual do Sudoeste da Bahia,
Itapetinga (BA), 45700-000, Brazil}

\author{Edilberto O. Silva\orcidlink{0000-0002-0297-5747}}
\email{edilberto.silva@ufma.br}
\affiliation{Programa de P\'os-Gradua\c c\~ao em F\'{\i}sica \& Coordena\c c\~ao do Curso de F\'{\i}sica -- Bacharelado, Universidade Federal do Maranh\~{a}o, 65085-580 S\~{a}o Lu\'{\i}s, Maranh\~{a}o, Brazil}

\begin{abstract}
Temporal sparsity compares the mean interval between emitted quanta with a
characteristic single-quantum timescale. We study this operational quantity for
the rotating draining-bathtub acoustic horizon in two spatial dimensions. The
Hawking temperature is fixed by the radial drain and does not depend on
circulation, while the geometric capture width and the transmission spectrum
do. A geometric black-body benchmark therefore predicts reduced temporal
separation as circulation increases. We then solve the radial scattering problem
and compute converged greybody-corrected number and energy fluxes, including the
superradiant sector. At a circulation-to-drain ratio of unity, the thermal-clock
sparsity is 0.44 whereas the spectrum-based sparsity is 1.82. The former crosses
the temporal-overlap benchmark near a ratio of 0.40, while the latter remains
above unity at all sampled circulation ratios. We also show that the binned sparsity
definition is logarithmically infrared sensitive in two spatial dimensions.
Rotation therefore enhances the number flux while hardening the spectrum, making
temporal overlap intrinsically dependent on the operational clock used to
resolve the emitted quanta.
\end{abstract}

\maketitle

% ===============================================================
\section{Introduction}
% ===============================================================

Hawking's prediction that black holes emit particles with a temperature fixed by
surface gravity is one of the central results of quantum field theory in curved
spacetime~\cite{Hawking1974,Hawking1975}. Direct observation from astrophysical
black holes is exceptionally difficult, which motivated the analogue-gravity
programme initiated by Unruh: linear perturbations of suitable moving media can
propagate on effective geometries with horizons and reproduce the kinematics
behind Hawking emission~\cite{Unruh1981,Visser1998,BarceloLiberatiVisser}. A key
conceptual question is whether the effect survives the breakdown of a
relativistic dispersion relation at short wavelength. Studies of ultrashort
scales and modified dispersion have shown both the robustness of the Hawking
mechanism under broad conditions and the circumstances in which ultraviolet
physics can alter the spectrum~\cite{Jacobson1991,Unruh1995,CorleyJacobson1996,
UnruhSchutzhold2005,MacherParentani2009D,CoutantParentaniFinazzi2012,
FinazziParentani2012}. Analogue systems are therefore useful not merely because
they imitate a metric, but because they provide controlled settings in which
horizons, dispersion, mode conversion, and detector response can be disentangled.

Bose--Einstein condensates provide the most developed quantum realization of
this programme. Sonic horizons were proposed and analysed microscopically in
dilute condensates~\cite{Garay2000,Garay2001}, and density correlations were
identified as a characteristic signature of pair creation across an acoustic
horizon~\cite{Balbinot2008,Carusotto2008,MacherParentani2009A,Recati2009}.
Laboratory work subsequently realized sonic horizons~\cite{Lahav2010}, explored
black-hole-laser amplification~\cite{Steinhauer2014}, and reported spontaneous
Hawking correlations and entanglement~\cite{Steinhauer2016}. Later experiments
measured a thermal spectrum and its temperature~\cite{Munoz2019} and tested the
stationarity of the spontaneous emission~\cite{Kolobov2021}; complementary
analyses have examined quantum signatures, deviations from thermality, and the
time development of the correlations~\cite{Boiron2015,Isoard2020,
FabbriBalbinot2021}. These results make clear that a realistic characterization
of analogue Hawking emission requires more than a temperature alone: frequency
dependent transmission and the operational definition of the measured flux are
part of the physics.

Surface-wave systems provide a complementary classical arena in which dispersion
and rotational scattering are especially accessible. Gravity-wave analogues of
horizons were developed theoretically~\cite{SchutzholdUnruh2002}, followed by
observations of negative-frequency mode conversion and detailed studies of
horizon effects in moving water~\cite{Rousseaux2008,Rousseaux2010}. Stimulated
Hawking conversion was measured in an open-channel experiment~\cite{Weinfurtner2011},
while subsequent work clarified when shallow-water spectra are or are not close
to thermal~\cite{MichelParentani2014} and demonstrated scattering on a
transcritical analogue black hole~\cite{Euve2020}. Rotation adds another layer.
The draining-bathtub vortex, the canonical rotating acoustic black hole in two
spatial dimensions, supports an ergoregion, superradiant amplification,
quasinormal ringing, absorption, and characteristic ray structure
~\cite{BasakMajumdar2003,BertiCardosoLemos2004,CardosoLemosYoshida2004,
Oliveira2010,DolanOliveira2013,ChurilovStepanyants2019}. Superradiance has been
observed in a draining vortex~\cite{Torres2017}, and dispersive corrections to
that process have been analysed beyond the nondispersive effective metric
~\cite{PatrickWeinfurtner2020,Patrick2022}. The ray-tracing observables used
below, including the asymmetric critical impact parameters, were recently
worked out for the same rotating acoustic geometry~\cite{bathtub}. Related
gravitational black-hole analyses likewise show that thermodynamic quantities
and critical null-orbit observables can carry complementary information about
the same background~\cite{SilvaReisAhmed2026}.

A separate feature of Hawking emission is its temporal \emph{sparsity}. Exact
black-hole flux calculations already require greybody transmission factors
~\cite{Page1976I,Page1976II}, but the sparsity programme asks a different
operational question: how does the mean waiting time between quanta compare with
a timescale associated with an individual emitted quantum? Gray, Schuster,
Van-Brunt, and Visser showed that the Hawking cascade of a Schwarzschild black
hole is extremely dilute according to several natural clock conventions
~\cite{Gray2016}. Rotation can qualitatively modify that conclusion for Kerr
emission because superradiant channels increase the rate and alter the spectrum
~\cite{Hod2015}; higher-dimensional and massive-particle extensions further show
that sparsity is sensitive to dimensionality and to the definition of the
single-quantum timescale~\cite{Schuster2019,SchusterThesis}. Recent
applications to deformed black-hole backgrounds have also used Hawking sparsity
as an independent radiative diagnostic alongside conventional thermodynamic
quantities~\cite{AhmedAlBadawiSilva2026}. This makes a rotating acoustic
horizon particularly useful: its Hawking temperature is fixed
by the radial drain, whereas circulation changes capture and transmission. The
thermal scale and the rotational scattering scale can therefore be varied
without being conflated.

In this work we formulate that problem explicitly for the rotating draining
bathtub. We first reconstruct the black-body sparsity benchmark in two spatial
dimensions and show how the geometric capture width produces a simple
circulation dependence. We then solve the partial-wave scattering problem with
explicit convergence tests and compute independent number- and energy-weighted
greybody lengths. This leads naturally to two distinct overlap diagnostics: a
fixed thermal-clock measure and a spectrum-based measure built from the actual
mean emitted frequency. We also show that the binned definition is
logarithmically infrared sensitive in two spatial dimensions, so a detector or
observation-time scale is unavoidable. Superradiant modes are treated
explicitly and are also removed as a diagnostic to test the robustness of the
conclusions. Section~\ref{sec:geometry} introduces the rotating geometry,
Sec.~\ref{sec:sparsity2d} develops the two-dimensional sparsity measures, and
Sec.~\ref{sec:absorption} presents the greybody calculation and emission
results. A phenomenological bandwidth illustration and the corrected
shallow-water dispersion expansion are collected in Appendix~\ref{app:bandwidth}
so that they remain clearly separated from the central nondispersive result.

% ===============================================================
% ===============================================================
\section{Rotating acoustic geometry}
\label{sec:geometry}
% ===============================================================

We work with the $(2+1)$-dimensional rotating draining-bathtub acoustic black
hole in the form used in Refs.~\cite{BertiCardosoLemos2004,Oliveira2010,DolanOliveira2013,bathtub}. Throughout, we use natural acoustic units
$c_s=\hbar=k_B=1$. In laboratory-type polar coordinates the effective metric is
\begin{equation}
 ds^2=dt^2-\left(dr-\frac{A}{r}dt\right)^{\!2}
 -\left(r\,d\phi-\frac{B}{r}dt\right)^{\!2},
 \label{bathtubmetric}
\end{equation}
where $A$ controls the radial draining flow and $B$ the circulation. In
Boyer--Lindquist-like coordinates this becomes
\begin{equation}
 ds^2=g(r)\,dt^2-\frac{dr^2}{f(r)}+2B\,dt\,d\phi-r^2d\phi^2,
 \label{blform}
\end{equation}
with
\begin{equation}
 f(r)=1-\frac{A^2}{r^2},\qquad g(r)=1-\frac{A^2+B^2}{r^2}.
 \label{fg}
\end{equation}
The sonic horizon sits at $f(r_h)=0$ and the ergosurface at $g(r_e)=0$:
\begin{equation}
 r_h=A,\qquad r_e=\sqrt{A^2+B^2}.
 \label{horizons}
\end{equation}
Throughout we use the dimensionless circulation
\begin{equation}
 \beta\equiv\frac{B}{A},\qquad\text{so that}\qquad
 \frac{r_e}{r_h}=\sqrt{1+\beta^2}.
 \label{beta}
\end{equation}

The acoustic surface gravity follows from the normal-flow profile,
$\kappa=\tfrac12\left|\partial_r\!\left(c_s^2-v_r^2\right)\right|_{r_h}$ with
$v_r=-A/r$ and $c_s=1$, so that $c_s^2-v_r^2=f(r)$ and
\begin{equation}
 \kappa=\frac{1}{A},\qquad
 T_H=\frac{\kappa}{2\pi}=\frac{1}{2\pi A}.
 \label{TH}
\end{equation}
This is the first ingredient of the argument: \emph{$T_H$ is independent of
$B$}. The horizon radius and surface gravity are set by the radial drain, while
circulation leaves the Hawking temperature unchanged. This is a structural
difference from Kerr, where the temperature depends on the spin, and it means
that any circulation dependence of the emission arises from the geometry and
mode structure rather than from a change in $T_H$.

Because the geometry is rotating, we also confirm \eqref{TH} covariantly. The horizon generator is
$\chi=\partial_t+\Omega_H\partial_\phi$ with
$\Omega_H=-g_{t\phi}/g_{\phi\phi}\big|_{r_h}=B/A^2$, and one verifies directly
that $\chi^2$ vanishes at $r=A$. Evaluating
$\kappa^2=-\tfrac12(\nabla^\mu\chi^\nu)(\nabla_\mu\chi_\nu)$ on the horizon
gives $\kappa^2=1/A^2$, so that the Killing-vector definition and the
normal-flow definition agree identically, and neither acquires a dependence on
$B$. The circulation enters $\chi$ through $\Omega_H$ but cancels from $\kappa$.

The second ingredient is the capture width. For a null ray with conserved
energy $E$, angular momentum $L$, and impact parameter $b=L/E$, the inverse
metric gives the radial equation
\begin{equation}
 \frac{\dot r^{\,2}}{E^2}=\mathcal R(r;b)
 =1-\frac{2Bb}{r^2}-\frac{b^2(r^2-A^2-B^2)}{r^4}.
 \label{rayradial}
\end{equation}
The capture boundary is an unstable circular null orbit, hence a double root of
$\mathcal R$: $\mathcal R(r_c;b_c)=0$ and
$\partial_r\mathcal R(r_c;b_c)=0$. Eliminating $r_c$ yields the two critical
impact parameters
\begin{equation}
 \frac{b_c^{\pm}}{A}=-2\beta\pm2\sqrt{1+\beta^2},
 \label{bc}
\end{equation}
in agreement with Ref.~\cite{bathtub}. Their midpoint and width are
\begin{equation}
 b_{\rm mid}=-2A\beta,\qquad
 \Delta b\equiv b_c^{+}-b_c^{-}=4A\sqrt{1+\beta^2}.
 \label{shadow}
\end{equation}
The centroid shift is linear in $\beta$ and the width grows as
$\sqrt{1+\beta^{2}}$; both are exact and independent of any source
model~\cite{bathtub}. Comparing \eqref{beta} and \eqref{shadow},
\begin{equation}
 \frac{\Delta b}{4A}=\frac{r_e}{r_h},
 \label{coincidence}
\end{equation}
that is, the capture width and the ergosurface radius scale identically. This
coincidence produces the compact result of the next section.

% ===============================================================
\section{Sparsity in \texorpdfstring{$(2+1)$}{(2+1)} dimensions}
\label{sec:sparsity2d}
% ===============================================================

The sparsity measures of Ref.~\cite{Gray2016} compare a mean waiting time
between emitted quanta with a timescale associated with a typical quantum. In
$d$ spatial dimensions the number flux from a black body with
$(d-1)$-dimensional boundary measure $\Sigma$ scales as $\Sigma T^d$. Thus the
familiar $A_{\rm eff}T^3$ law is specific to $d=3$; for $d=2$ the emitting
measure is a length and the flux is quadratic in temperature.

By detailed balance, for an absorber of capture width $\Sigma$ immersed in an
isotropic two-dimensional bath of massless bosons,
\begin{equation}
 \dot N=\int\!\frac{d^2k}{(2\pi)^2}\,\frac{\Sigma}{e^{k/T}-1}
 =\frac{\Sigma T^2\zeta(2)}{2\pi}
 =\frac{\pi\Sigma T^2}{12}.
 \label{flux2d}
\end{equation}
This agrees with the eikonal mode sum
$\dot N=\sum_m\int(d\omega/2\pi)\,\Gamma_m n_B$ when
$\sum_m\Gamma_m\rightarrow\omega\Sigma$. The mean gap is
$\tau_{\rm gap}=1/\dot N$.

For the fiducial single-quantum clock we use one oscillation period evaluated at
the mean emitted frequency,
\begin{equation}
 \langle\omega\rangle=T\,
 \frac{\int_0^\infty x^2dx/(e^x-1)}
 {\int_0^\infty x\,dx/(e^x-1)}
 =\frac{2\zeta(3)}{\zeta(2)}\,T,
 \label{meanomega}
\end{equation}
and define
\begin{equation}
 \eta\equiv\frac{\tau_{\rm gap}}{2\pi/\langle\omega\rangle}
 =\frac{12\zeta(3)}{\pi^2\zeta(2)}\,\frac{1}{\Sigma T}.
 \label{eta2d}
\end{equation}
The dependence is linear, rather than quadratic, in the ratio of wavelength to
emitting size. Throughout this paper $\eta\lesssim1$ means that temporal overlap
is favoured according to the specified clock; it does not by itself establish a
classical coherent field.

Using $T=T_H$ from Eq.~\eqref{TH} and $\Sigma=\Delta b$ from
Eq.~\eqref{shadow} gives the geometric capture-width benchmark
\begin{equation}
 \eta_{\rm geo}(\beta)=
 \frac{36\zeta(3)}{\pi^3\sqrt{1+\beta^2}}
 =\eta_{\rm geo}(0)\frac{r_h}{r_e},
 \label{result}
\end{equation}
with $\eta_{\rm geo}(0)=36\zeta(3)/\pi^3\simeq1.396$. The compact
$\beta$ dependence follows exactly from the geometry \emph{within the
capture-width black-body approximation}: $T_H$ remains fixed while the capture
width grows. It should not be read as the exact Hawking flux of the rotating
system, because the latter also contains the chemical-potential-like combination
$\omega-m\Omega_H$ and frequency-dependent greybody factors.

\begin{figure*}[tbhp]
\centering
\includegraphics[width=\textwidth]{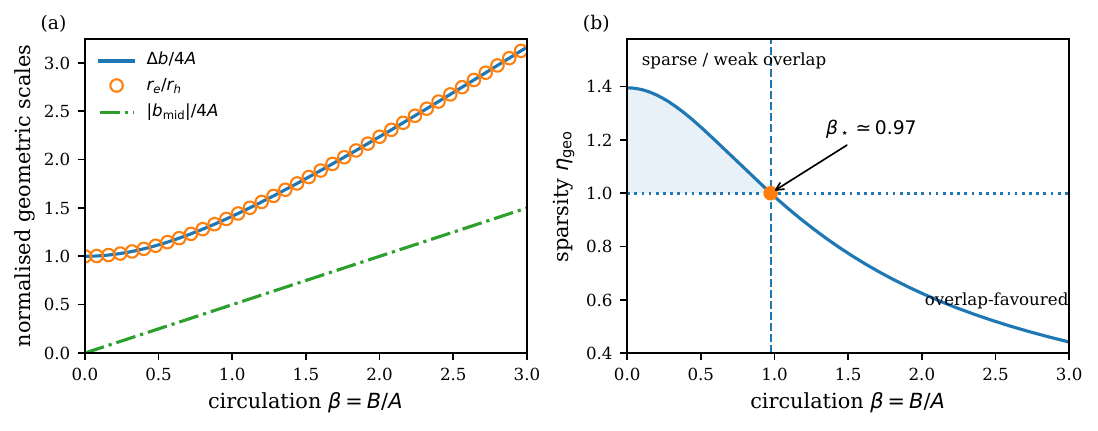}
\caption{(a) Geometric scales of the rotating acoustic horizon. The normalised
capture width $\Delta b/4A$ and ergosurface radius $r_e/r_h$ coincide,
Eq.~\eqref{coincidence}, while the shadow centroid grows linearly.
(b) Geometric capture-width sparsity, Eq.~\eqref{result}. The line $\eta=1$
is an operational temporal-overlap benchmark, not a quantum-to-classical phase
boundary; the geometric crossing occurs at $\beta\simeq0.97$.}
\label{fig:circ}
\end{figure*}

The geometric benchmark crosses $\eta=1$ at
\begin{equation}
 \beta_\star^{({\rm geo})}
 =\sqrt{\left(\frac{36\zeta(3)}{\pi^3}\right)^2-1}
 \simeq0.97.
 \label{threshold}
\end{equation}
Figure~\ref{fig:circ} displays the geometry and this overlap benchmark.

\subsection{Clock convention}

The ratio in Eq.~\eqref{result} is fixed by the geometric benchmark, but its
absolute normalisation depends on the single-quantum clock. Table~\ref{tab:conventions}
lists four transparent choices. The first three use a complete oscillation
period. Because the ideal $(2+1)$ number spectrum is largest at $\omega=0$,
there is no nonzero number-spectrum Wien peak; the peak entry therefore refers
explicitly to the \emph{energy} spectrum.

\begin{table}[tbhp]
\caption{Geometric sparsity at zero circulation and the associated
$\eta=1$ overlap crossing for four clock conventions. A dash indicates that the
benchmark is already below unity at $\beta=0$.}
\label{tab:conventions}
\centering
\begin{tabular}{lcc}
\toprule
$\tau_{\rm clock}$ & $\eta(0)$ & $\beta_\star$ \\
\midrule
$2\pi/T_H$                                      & 0.955 & --- \\
$2\pi/\langle\omega\rangle$ (fiducial)          & 1.396 & 0.97 \\
$2\pi/\omega_{\rm peak}^{(E)}$                  & 1.522 & 1.15 \\
$1/T_H$                                         & 6.00  & 5.92 \\
\bottomrule
\end{tabular}
\end{table}

The period-based conventions place the nonrotating benchmark near unity, so the
side of the overlap line depends on the clock. We use
$2\pi/\langle\omega\rangle$ because it is tied to the emitted quanta. The
$1/T_H$ entry is shown only to expose the $2\pi$ convention dependence; it is
not one complete oscillation period. For comparison, the corresponding geometric-optics black-body estimate for a
scalar Schwarzschild black hole, evaluated with the same average-frequency
clock, is $\eta_{\rm Sch}=26.285$~\cite{Gray2016}, about nineteen times the
present nonrotating geometric value. The often-quoted number
$64\pi^3/27\simeq73.5$ in Ref.~\cite{Gray2016} is the dimensionless ratio
$\lambda_{\rm thermal}^2/A_{\rm effective}$, not itself the average-frequency
sparsity.

\subsection{Infrared sensitivity of a binned measure}

Gray et al. also introduced a binned measure that weights each spectral bin by
its own oscillation period~\cite{Gray2016}. In the present notation,
\begin{equation}
 \eta_{\rm bin}^{-1}=
 \int_0^\infty\frac{2\pi}{\omega}
 \frac{d^2N}{dt\,d\omega}\,d\omega.
 \label{etabin}
\end{equation}
For the geometric $(2+1)$ spectrum,
\begin{equation}
 \frac{d^2N}{dt\,d\omega}
 =\frac{\Sigma}{2\pi}\frac{\omega}{e^{\omega/T}-1},
 \label{numbergeo}
\end{equation}
so Eq.~\eqref{etabin} contains
$\int_0^\infty d\omega/(e^{\omega/T}-1)$ and diverges logarithmically in the
infrared. Introducing an operational low-frequency resolution
$\omega_{\min}$ gives
\begin{equation}
 \eta_{\rm bin}(\omega_{\min})=
 \left\{\Sigma T\left[-\ln\!\left(1-e^{-\omega_{\min}/T}\right)\right]\right\}^{-1}.
 \label{etabincut}
\end{equation}
The divergence is not cured by the exact low-frequency greybody factor. As
shown below by the absorption limit in Eq.~\eqref{lowabs}, the $m=0$ channel
dominates as $\omega\to0$ and obeys
$\Gamma_0\simeq2\pi A\,\omega$; the emitted number spectrum therefore again
approaches a constant. Thus a binned sparsity in two
spatial dimensions necessarily depends on a finite observation time, detector
bandwidth, finite system size, or another infrared scale. This does not
invalidate the mean-frequency measure in Eq.~\eqref{eta2d}; it shows that
``sparsity'' is more explicitly operational in $(2+1)$ dimensions than in the
usual $(3+1)$ black-body benchmark.

% ===============================================================
\section{Absorption and Hawking emission}
\label{sec:absorption}
% ===============================================================

The geometric estimate isolates the capture-width dependence, but it does not
determine how efficiently each frequency and azimuthal mode reaches infinity.
That information is contained in the absorption problem. By detailed balance,
the same transmission coefficient that measures classical absorption of an
incident wave supplies the greybody factor in the Hawking spectrum.

\subsection{Radial equation and absorption length}

Let $\Phi$ be a minimally coupled massless phonon satisfying $\Box_g\Phi=0$.
With
\begin{equation}
 \Phi(t,r,\phi)=e^{-i\omega t+im\phi}\frac{u_{\omega m}(r)}{\sqrt r},
 \qquad \frac{dr_*}{dr}=\frac{1}{f(r)},
 \label{separation}
\end{equation}
the tortoise coordinate is
$r_*=r+(A/2)\ln[(r-A)/(r+A)]$, up to an additive constant, and the radial
equation becomes
\begin{equation}
 \frac{d^2u_{\omega m}}{dr_*^2}
 +\left[\left(\omega-\frac{mB}{r^2}\right)^2-V_m(r)\right]u_{\omega m}=0,
 \label{radialeq}
\end{equation}
where
\begin{equation}
 V_m(r)=f(r)\left[\frac{m^2-1/4}{r^2}+\frac{5A^2}{4r^4}\right].
 \label{potential}
\end{equation}
At the horizon the relevant frequency is
$\widetilde\omega=\omega-m\Omega_H$, while at infinity the solutions are free
waves of frequency $\omega$.

For a unit wave incident from infinity,
\begin{align}
 u_{\omega m}&\sim e^{-i\omega r_*}+{\cal R}_m e^{+i\omega r_*},
 &&r_*\to+\infty,\nonumber\\
 u_{\omega m}&\sim {\cal T}_m e^{-i\widetilde\omega r_*},
 &&r_*\to-\infty.
 \label{boundary}
\end{align}
Wronskian conservation gives
\begin{equation}
 \Gamma_m(\omega)\equiv1-|{\cal R}_m|^2
 =\frac{\omega-m\Omega_H}{\omega}|{\cal T}_m|^2.
 \label{wronskian}
\end{equation}
The partial and total absorption lengths are
\begin{equation}
 \sigma_m(\omega)=\frac{\Gamma_m(\omega)}{\omega},\qquad
 \sigma_{\rm abs}(\omega)=\sum_{m=-\infty}^{\infty}\sigma_m(\omega),
 \label{absorptionlength}
\end{equation}
consistent with Refs.~\cite{BasakMajumdar2003,Oliveira2010,DolanOliveira2013}. In a co-rotating superradiant interval
$0<\omega<m\Omega_H$, $\Gamma_m<0$: the reflected classical wave carries more
flux than the incident wave.

Two analytic limits provide stringent checks,
\begin{equation}
 \lim_{\omega A\to0}\sigma_{\rm abs}=2\pi A,
 \label{lowabs}
\end{equation}
and
\begin{align}
 \lim_{\omega A\to\infty}\sigma_{\rm abs}&=\Delta b,
 \nonumber\\
 \sum_m\Gamma_m&\longrightarrow\omega\Delta b.
 \label{highabs}
\end{align}
These are the horizon circumference and the geometric capture width,
respectively~\cite{Oliveira2010,DolanOliveira2013}.

\subsection{Number and energy spectra}

For bosonic phonons the differential fluxes measured at infinity are
\begin{align}
 \frac{d^2N}{dt\,d\omega}
 &=\frac{1}{2\pi}\sum_m
 \frac{\Gamma_m(\omega)}{e^{(\omega-m\Omega_H)/T_H}-1},
 \label{numberspectrum}\\
 \frac{d^2E}{dt\,d\omega}
 &=\frac{\omega}{2\pi}\sum_m
 \frac{\Gamma_m(\omega)}{e^{(\omega-m\Omega_H)/T_H}-1}.
 \label{energyspectrum}
\end{align}
Both numerator and denominator change sign in the superradiant interval, so
the emitted number and energy flux densities remain positive. Numerically we evaluate this
ratio in a cancellation-safe form using Eq.~\eqref{wronskian}, which also
removes the apparent $0/0$ at $\omega=m\Omega_H$.

In the nonrotating geometric-optics approximation,
$\sum_m\Gamma_m=\omega\Sigma$ with $\Sigma=\Delta b$, giving
\begin{equation}
 \dot N_{\rm geo}=\frac{\zeta(2)}{2\pi}\Sigma T_H^2,
 \qquad
 \dot E_{\rm geo}=\frac{\zeta(3)}{\pi}\Sigma T_H^3,
 \label{geofluxes}
\end{equation}
and therefore
\begin{equation}
 \langle\omega\rangle_{\rm geo}
 =\frac{2\zeta(3)}{\zeta(2)}T_H.
 \label{geomean}
\end{equation}
The \emph{energy} spectrum peaks at
$\omega_{\rm peak}^{(E)}/T_H\simeq1.594$. Dimensional analysis gives
$\dot N\propto A^{-1}$ and $\dot E\propto A^{-2}$, with dimensionless profiles
depending only on $\beta$.

\subsection{Numerical implementation and convergence}

All reported greybody calculations use the scale choice $A=1$; the scaling with
$A$ is restored analytically. We start at $r_{\rm in}=A(1+10^{-5})$ with the
ingoing horizon solution and integrate Eq.~\eqref{radialeq} with an eighth-order
Dormand--Prince method (DOP853), relative tolerance $5\times10^{-7}$ and
absolute tolerance $5\times10^{-9}$. The asymptotic matching radius is chosen
as
\begin{equation}
 r_{\max}/A=\max\left(80,\frac{30}{\omega A}\right).
 \label{rmax}
\end{equation}
The frequency integral is split at the superradiant boundary when present and
extended to $\omega_{\max}A=\max(3,m\beta+2)$ for $m>0$; the remaining thermal
tail is exponentially negligible at the quoted precision. A composite Simpson
rule is used on a denser grid around $\omega=m\Omega_H$.

The mode sum is converged independently for number and energy. At $\beta=1$ we
retain $-2\le m\le12$; the $m=12$ term changes the total number and energy
integrals by approximately $4\times10^{-7}$ and $5\times10^{-6}$,
respectively. The counter-rotating tail is already negligible at the lower
limit: the number integrals for $m=-2,-3,-4$ are approximately
$4.8\times10^{-8}$, $2.5\times10^{-11}$, and $1.3\times10^{-14}$. Doubling
the frequency grid changes the dominant $m=0,\ldots,4$ mode integrals by less
than $5\times10^{-5}$ fractionally. Moving the starting point from
$r/A-1=10^{-5}$ to $10^{-6}$ changes a representative $m=1$, $\omega A=0.8$
greybody factor by less than $8\times10^{-5}$, while extending the positive-$m$
thermal tail by $1/A$ changes the corresponding number and energy integrals by
less than $7\times10^{-6}$ and $2\times10^{-5}$. The Wronskian residual stays
below $2\times10^{-5}$ on the production grids. The reproducibility package
records these tests and the cumulative finite-$m_{\max}$ differences relative
to the $m_{\max}=12$ production sum. This explicit energy-flux test is
important: truncating at $m\le2$ is already adequate-looking for some
number-flux diagnostics but underestimates $\dot E$ severely at $\beta=1$.

We encode greybody corrections in separate number- and energy-weighted effective
lengths,
\begin{align}
 \Sigma_N&=\frac{1}{\zeta(2)T_H^2}\sum_m\int_0^\infty d\omega\,
 \frac{\Gamma_m}{e^{(\omega-m\Omega_H)/T_H}-1},\nonumber\\
 \Sigma_E&=\frac{1}{2\zeta(3)T_H^3}\sum_m\int_0^\infty d\omega\,
 \frac{\omega\Gamma_m}{e^{(\omega-m\Omega_H)/T_H}-1}.
 \label{effectivelengths}
\end{align}
Then $\dot N=\pi\Sigma_NT_H^2/12$ and
$\dot E=\zeta(3)\Sigma_ET_H^3/\pi$. Rotation therefore cannot in general be
represented by one effective emitting length: $\Sigma_E\ne\Sigma_N$.

Two clocks make this distinction explicit. The fixed thermal-clock quantity is
\begin{equation}
 \eta_T\equiv\frac{\langle\omega\rangle_{\rm geo}}{2\pi\dot N},
 \label{etathermal}
\end{equation}
whereas the spectrum-based overlap measure is
\begin{equation}
 \eta_{\rm sp}\equiv
 \frac{\langle\omega\rangle_{\rm em}}{2\pi\dot N}
 =\frac{\dot E}{2\pi\dot N^2},\qquad
 \langle\omega\rangle_{\rm em}=\frac{\dot E}{\dot N}.
 \label{etaspectral}
\end{equation}

\begin{table}[!t]
\caption{Converged absorption-weighted emission for $A=1$. The effective
lengths are defined in Eq.~\eqref{effectivelengths}. The thermal clock uses
Eq.~\eqref{geomean}; $\eta_{\rm sp}$ uses the actual mean emitted frequency.}
\label{tab:greybody}
\centering
\begin{tabular}{cccccc}
\toprule
$\beta$ & $\Sigma_N/\Delta b$ & $\Sigma_E/\Delta b$
& $\langle\omega\rangle_{\rm em}/T_H$ & $\eta_T$ & $\eta_{\rm sp}$ \\
\midrule
$0.0$ & 1.14 & 1.04 & 1.33 & 1.23 & 1.12 \\
$0.5$ & 1.39 & 2.33 & 2.44 & 0.90 & 1.50 \\
$1.0$ & 2.25 & 9.33 & 6.06 & 0.44 & 1.82 \\
\bottomrule
\end{tabular}
\end{table}

The low-frequency enhancement is already visible at $\beta=0$, where the
horizon-circumference limit $2\pi A$ exceeds the geometric width $4A$.
Increasing circulation activates co-rotating channels and raises $\Sigma_N$,
so the full greybody thermal-clock measure crosses unity near
\begin{equation}
 \beta_\star^{(T)}\simeq0.40.
 \label{betastargrey}
\end{equation}
The energy-weighted length grows much more rapidly: at $\beta=1$ the converged
value is $\Sigma_E/\Delta b\simeq9.33$ and the mean emitted frequency reaches
$6.06T_H$. Consequently $\eta_{\rm sp}$ remains above unity at all sampled values in the
range and in fact rises from $1.12$ at zero circulation to $1.82$ at
$\beta=1$. The emission becomes less sparse with respect to a fixed thermal
clock while simultaneously hardening spectrally.

\begin{figure*}[tbhp]
\centering
\includegraphics[width=\textwidth]{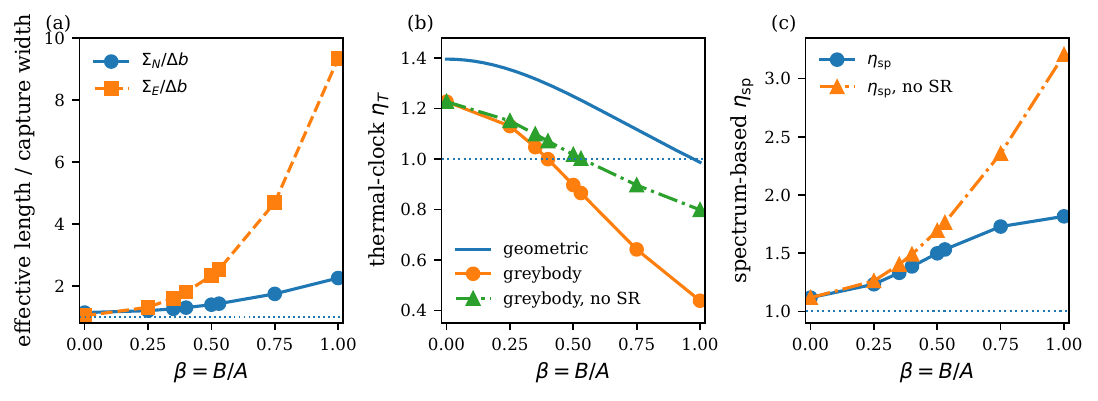}
\caption{Greybody-corrected quantities as functions of circulation. (a)
Number- and energy-weighted effective lengths, showing the much faster growth of
$\Sigma_E$. (b) Thermal-clock sparsity: geometric benchmark, full greybody
result, and greybody result with the superradiant interval removed. The full
crossing is near $\beta\simeq0.40$ and the non-superradiant crossing near
$0.53$. (c) Spectrum-based sparsity $\eta_{\rm sp}$, shown both with and without
the superradiant interval. It remains above the overlap line $\eta=1$ at all sampled values for
$0\le\beta\le1$. Points are direct numerical calculations; connecting curves
are guides to the eye.}
\label{fig:greybody}
\end{figure*}

\subsection{Superradiant sector and mode split}

For $\beta\ne0$ the geometry has an ergoregion and supports superradiant
amplification~\cite{BasakMajumdar2003,Torres2017,PatrickWeinfurtner2020}. Ref.~\cite{Gray2016} emphasises
that rotating systems require an explicit convention for superradiant modes,
because quantum vacuum radiation in that interval need not be grouped with the
non-superradiant Hawking flux for every operational definition.

At eikonal level a mode with impact parameter $b=m/\omega$ is captured only if
$b\le b_c^+$, hence
\begin{equation}
 \frac{m\Omega_H}{\omega}\le b_c^+\Omega_H
 =2\beta\left(\sqrt{1+\beta^2}-\beta\right)
 \equiv\mathcal{G}(\beta).
 \label{superrad}
\end{equation}
The bound is particularly transparent after rationalizing the difference in
Eq.~\eqref{superrad}. For $\beta>0$,
\begin{align}
 \mathcal{G}(\beta)
 &=2\beta\left(\sqrt{1+\beta^2}-\beta\right)\nonumber\\
 &=\frac{2\beta}{\sqrt{1+\beta^2}+\beta}
 =\frac{2}{\sqrt{1+\beta^{-2}}+1}.
 \label{Grationalized}
\end{align}
The first rationalized form makes the finite-circulation bound explicit:
$\sqrt{1+\beta^2}>\beta$, so its denominator is strictly larger than
$2\beta$. The last form makes the large-circulation limit immediate. Including
$\mathcal{G}(0)=0$, one therefore has
\begin{equation}
 \begin{aligned}
  \mathcal{G}(0)&=0,\qquad
  0<\mathcal{G}(\beta)<1\quad(\beta>0),\\
  \lim_{\beta\to\infty}\mathcal{G}(\beta)&=1.
 \end{aligned}
 \label{Gbound}
\end{equation}
Thus, for a co-rotating mode with $m>0$, the superradiant condition
$0<\omega<m\Omega_H$ requires
$m\Omega_H/\omega>1$, whereas eikonal capture implies
$m\Omega_H/\omega\le\mathcal{G}(\beta)<1$ for every finite $\beta$.
Geometrically captured rays and the superradiant sector are therefore disjoint
in the strict eikonal limit, with the capture bound approaching the
superradiant threshold from below as $\beta\to\infty$. The laboratory
observation of Ref.~\cite{Torres2017} was correspondingly a scattering
experiment.

At finite wavelength the eikonal separation is not exact, so we also remove the
superradiant interval directly from the mode integrals as a diagnostic. In the
converged calculation, superradiant frequencies account for about $12\%$ of the
number flux at $\beta=0.5$ and $45\%$ at $\beta=1$. Excluding them moves the
thermal-clock $\eta_T=1$ crossing from $\beta\simeq0.40$ to
$\beta\simeq0.53$. The spectrum-based measure becomes even larger: after
removing the superradiant interval, $\eta_{\rm sp}\simeq1.69$ at $\beta=0.5$
and $\simeq3.20$ at $\beta=1$. Thus the conclusion that the spectrum-based sparsity $\eta_{\rm sp}$ remains
above unity at all sampled values for $0\le\beta\le1$ is not driven by the
inclusion of superradiant modes.

\section{Conclusions}

We have shown that temporal sparsity in a rotating acoustic horizon is not a
single geometry-only number. In two spatial dimensions the black-body number
flux scales with an emitting length times the square of the temperature, which
makes the geometric benchmark linear in the ratio of wavelength to capture
size. For the draining bathtub, circulation leaves the Hawking temperature
unchanged while increasing the capture width, so the geometric benchmark
falls monotonically with rotation.

The greybody calculation gives the more informative result. Number- and
energy-weighted effective lengths separate strongly as circulation increases,
because the same co-rotating channels that enhance the particle rate also harden
the emitted spectrum. With the mode sum converged through the positive-angular
momentum tail at unit circulation, the fixed thermal-clock sparsity is 0.44,
whereas the spectrum-based sparsity is 1.82. The former crosses the temporal
overlap line near a circulation-to-drain ratio of 0.40; the latter stays above
unity at all sampled circulation ratios. Removing superradiant frequencies shifts
the thermal-clock crossing to about 0.53 and increases the spectrum-based
sparsity, so this qualitative distinction is not an artefact of how the
superradiant sector is counted.

Two qualifications are essential. First, unity is an operational overlap
benchmark rather than a quantum-to-classical boundary. Second, the binned
measure is logarithmically infrared sensitive in two spatial dimensions and
therefore requires a finite observation time, detector bandwidth, system size,
or another low-frequency scale. The robust conclusion is consequently that
rotation enhances the emitted number flux relative to a fixed thermal clock but
simultaneously shifts weight to higher frequencies. In rotating analogue
Hawking emission, sparsity is therefore intrinsically clock dependent, and
number and energy fluxes must be analysed together. Appendix~\ref{app:bandwidth}
shows separately how finite bandwidth and medium dispersion can further modify
an operational measurement.

\section*{Acknowledgments}
Fernando M. Belchior acknowledges support from the Conselho Nacional de Desenvolvimento Cient\'{\i}fico e Tecnol\'ogico (CNPq), grant No.~151845/2025-5. E. O. Silva acknowledges the support from Conselho Nacional de
Desenvolvimento Cient\'{\i}fico e Tecnol\'{o}gico (CNPq) (grant 306308/2022-3),
Funda\c c\~ao de Amparo \`a Pesquisa e ao Desenvolvimento Cient\'{\i}fico e
Tecnol\'{o}gico do Maranh\~ao (FAPEMA) (grant UNIVERSAL-06395/22), and
Coordena\c c\~ao de Aperfei\c coamento de Pessoal de N\'{\i}vel Superior
(CAPES) -- Brazil (Finance Code 001). JAASR acknowledges partial financial support from UESB through Grant AuxPPI (Edital No. 267/2024) and also gratefully recognizes support from FAPESB–CNPq/Produtividade (Grant No. 12243/2025, TOB-BOL2798/2025).

\appendix
\section{Dispersion and finite-bandwidth illustration}
\label{app:bandwidth}

The effective-metric description of a real analogue medium fails at sufficiently
short wavelength, but the physical modification is a dispersion relation in
wave number, not in general a hard cutoff in laboratory frequency. For a dilute
Bose--Einstein condensate the local comoving Bogoliubov dispersion may be written
schematically as~\cite{MacherParentani2009A,Recati2009,FinazziParentani2012}
\begin{equation}
 \omega_{\rm com}^2=c_s^2k^2\left(1+\frac{\xi^2k^2}{2}\right),
 \label{bogoliubov}
\end{equation}
with $\xi$ the healing length. This branch is superluminal at large $k$.

Surface waves in water behave differently at the first dispersive order~\cite{SchutzholdUnruh2002,Rousseaux2008,Rousseaux2010,MichelParentani2014}. Their
local gravity--capillary dispersion in the fluid-comoving frame is
\begin{equation}
 \omega_{\rm com}^2=\left(gk+\frac{\sigma k^3}{\rho}\right)\tanh(kh).
 \label{fulldispwater}
\end{equation}
Writing $\ell_c=\sqrt{\sigma/(\rho g)}$ and using
$\tanh(kh)=kh-(kh)^3/3+\cdots$ gives
\begin{equation}
 \omega_{\rm com}^2=gh\,k^2\left[1+
 \left(\ell_c^2-\frac{h^2}{3}\right)k^2+O(k^4)\right].
 \label{waterdispersion}
\end{equation}
For water $\ell_c\simeq2.7\,$mm. In the vortex experiment of
Ref.~\cite{Torres2017}, $h=6.25\,$cm, so $h^2/3\gg\ell_c^2$ and the leading
finite-depth correction is strongly \emph{subluminal}. At still shorter
wavelengths capillarity bends the full branch upward, but Eq.~\eqref{fulldispwater}
should then be used directly. Thus shallow-water dispersion is not obtained by
simply replacing the Bogoliubov healing length with a capillary length.

\begin{figure*}[t]
\centering
\includegraphics[width=\textwidth]{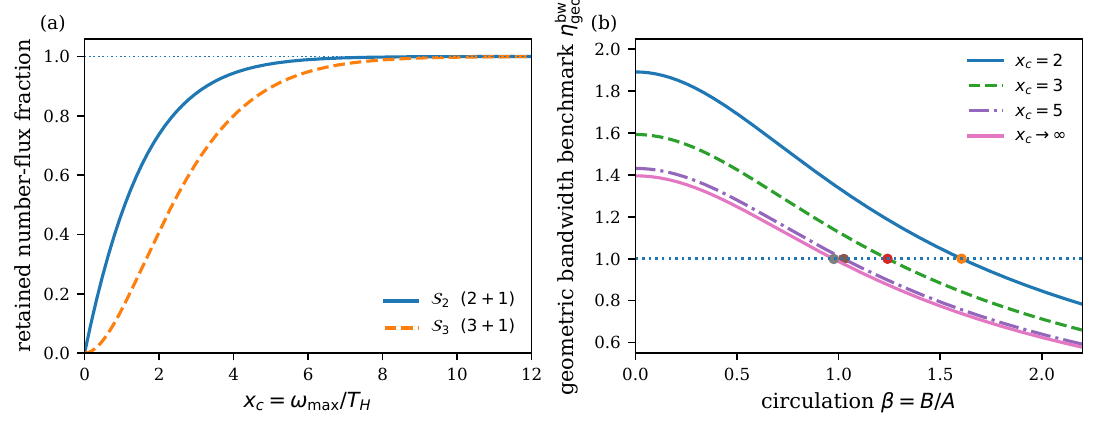}
\caption{Phenomenological bandwidth illustration. (a) Fraction of the ideal
number flux retained below a hard frequency cutoff in two and three spatial
dimensions. No experimental ``lab regime'' is assigned because $x_c$ must be
inferred from the actual dispersion and detector response. (b) Bandwidth-modified geometric sparsity after applying the $(2+1)$ cutoff. The line $\eta=1$ is a temporal-overlap benchmark.}
\label{fig:disp}
\end{figure*}

To isolate one generic consequence of finite experimental or model bandwidth,
we nevertheless introduce an \emph{illustrative phenomenological} hard cutoff
$\omega_{\max}$ and define $x_c=\omega_{\max}/T_H$. This is not claimed to be
the physical dispersion law of either a condensate or a water vortex. In the
geometric $(2+1)$ number flux, the surviving fraction is
\begin{equation}
 \mathcal{S}_2(x_c)=\frac{1}{\zeta(2)}
 \int_0^{x_c}\frac{x\,dx}{e^x-1},
 \label{suppression}
\end{equation}
so the bandwidth-modified geometric benchmark is
\begin{equation}
 \eta_{\rm geo}^{\rm bw}(\beta,x_c)=
 \frac{\eta_{\rm geo}(\beta)}{\mathcal{S}_2(x_c)}.
 \label{etageobw}
\end{equation} The corresponding
$(3+1)$ number-flux suppression contains $x^2$ and is stronger at the same
$x_c$. Figure~\ref{fig:disp} shows this dimensional comparison and the shift of
the geometric overlap crossing.

The resulting illustrative crossings are listed in Table~\ref{tab:threshold}.
They quantify sensitivity to a finite bandwidth, not a prediction for a
specific laboratory apparatus.

\begin{table}[tbhp]
\caption{Illustrative hard-bandwidth dependence of the geometric
overlap crossing. $\mathcal{S}_2$ is the retained number-flux fraction.}
\label{tab:threshold}
\centering
\begin{tabular}{ccc}
\toprule
$x_c$ & $\mathcal{S}_2(x_c)$ & $\beta_\star$ \\
\midrule
$2$      & 0.738 & 1.61 \\
$3$      & 0.876 & 1.24 \\
$5$      & 0.975 & 1.02 \\
$\infty$ & 1     & 0.97 \\
\bottomrule
\end{tabular}
\end{table}

A physical analysis of a specific experiment should instead propagate the full
dispersive mode equation and detector transfer function~\cite{PatrickWeinfurtner2020}. The hard-cutoff model
is retained only because it cleanly displays how any loss of high-frequency
number flux raises a thermal-clock sparsity estimate.

\FloatBarrier

\end{document}